\documentclass[11pt,a4paper]{article}
\pdfoutput=1

\usepackage{jheppub}
\usepackage{amsmath}

\graphicspath{ {fig/} }

\newcommand{\p}{p \hspace{-1.0ex} /}
\newcommand{\M}{ {\rm M} }
\newcommand{\e}{ {\rm e} }

\newcommand{\finite}{ {\rm finite} }
\newcommand{\C}{ \mathcal{C} }

\newcommand{\MS}{\overline{ \rm MS}}

\newcommand{\reduze}{{\tt Reduze}}
\newcommand{\hepforge}{{\tt HepForge}}
\newcommand{\qgraf}{{\tt Qgraf}}

\newcommand{\form}{{\tt Form}}
\newcommand{\jaxodraw}{{\tt JaxoDraw}}

\allowdisplaybreaks[3]

\makeatletter
\renewcommand\@fpheader{} 
\renewcommand\@journal{}
\makeatother

\title{
The two-loop helicity amplitudes for $gg \to V_1 V_2\to 4~\mathrm{leptons}$
}

\preprint{{MITP/15-012, TTP15-014}}

\author[a]{Andreas von Manteuffel,}
\author[b]{Lorenzo Tancredi}

  \affiliation[a]{
  PRISMA Cluster of Excellence, Institute of Physics,
  Johannes Gutenberg University,\\
  55099~Mainz, Germany}
  
  \affiliation[b]{
  Institut f\"{u}r Theoretische Teilchenphysik, Karlsruhe 
  Institute of Technology,\\ Engesserstrasse 7, 76128 Karlsruhe, Germany
  }

\emailAdd{manteuffel@uni-mainz.de}
\emailAdd{lorenzo.tancredi@kit.edu}


\abstract{We compute the two-loop massless QCD corrections to the helicity
amplitudes for the production of two electroweak gauge bosons in the
gluon fusion channel, $gg \to V_1 V_2$,
keeping the virtuality of the vector bosons $V_1$ and $V_2$ arbitrary
and taking their decays into leptons into account.
The amplitudes are expressed in terms of master integrals,
whose representation has been optimised for fast and 
reliable numerical evaluation.
We provide analytical results and a public {\tt C++} code for
their numerical evaluation on \hepforge\ at \url{http://vvamp.hepforge.org}.
}

\begin{document}
\unitlength1cm
\maketitle

\section{Introduction}
\label{sec:intro}

Pair production processes for electroweak vector bosons
provide a rich spectrum of observables, which are crucial
to test in depth the $SU(2)_L\times U(1)_Y$ gauge sector
of the Standard Model.
In particular, the production of pairs of resonant vector bosons
allows for precise studies of the electroweak triple gauge couplings, while
considering off-shell vector boson pairs
is required for precision Higgs phenomenology.
Furthermore, diboson production processes are important backgrounds
in direct new physics searches.
The main production channel for pairs of vector bosons at hadron colliders is
quark-antiquark annihilation and great progress has been achieved in the last years
with the computation of the next-to-next-to-leading order (NNLO) QCD corrections to 
$q \bar{q} \to \gamma \gamma$~\cite{Catani:2011qz}, $q \bar{q} \to Z \gamma$~\cite{Grazzini:2013bna}, 
$q \bar{q} \to ZZ$~\cite{Cascioli:2014yka} and $ q \bar{q} \to W^+W^-$~\cite{Gehrmann:2014fva} production at the LHC.
Furthermore, the fermionic NNLO corrections to $q \bar{q} \to \gamma^* \gamma^*$ were derived in~\cite{Anastasiou:2014nha}.

The gluon fusion channel contributes to
$\gamma \gamma$, $ZZ$, $Z \gamma$ and $W^+W^-$ production.
As a quark-loop induced process, its leading order (LO) cross section
is suppressed by two powers of $\alpha_s$
with respect to that of the quark channel.
This implies that it formally contributes only at NNLO in the perturbative
expansion of the hadronic process,
but numerical enhancements may be expected due to
the large gluon luminosities at typical energies
for diboson production at the LHC.
For the gluon-induced processes, the one-loop amplitudes and the corresponding
one-loop squared interference terms have been computed long ago~\cite{Costantini:1971cj,vanderBij:1988fb,Glover:1988rg,Glover:1988fe,Adamson:2002jb,Adamson:2002rm,Binoth:2006mf}.
Their impact on the total cross section was found
to range approximately from $5\%$ to more than $10\%$ for different final states at the LHC,
and to rise with increasing collider energy~\cite{Catani:2011qz,Grazzini:2013bna,Cascioli:2014yka,Gehrmann:2014fva}.
These values can substantially increase up to about $30\%$ when particular sets of 
cuts, relevant for example for Higgs boson searches, are applied~\cite{Binoth:2005ua, Cascioli:2013gfa}.
It is therefore clear that the inclusion of gluon channel contributions can be important in order
to achieve a description of the full process which matches the experimental precision.
Beyond the actual size of the known leading order corrections in the gluon channel,
it is unclear how large the associated theory uncertainty actually is.
By comparison with Higgs production in gluon fusion~\cite{Dawson:1990zj,Djouadi:1991tka,Spira:1995rr},
the conventional LO scale variation is not expected to allow for a reliable estimate
of the size of neglected higher order corrections.
The recent NNLO predictions for the total $ZZ$ and $W^+W^-$ production
cross sections take into account the quark channel at NNLO and the gluon channel
at LO, resulting in a scale uncertainty of about $3 \%$~\cite{Cascioli:2014yka,Gehrmann:2014fva}.
In order to thoroughly control the theory uncertainty to this level of precision,
it is therefore very desirable to compute
the next-to-leading order (NLO) contributions for the gluon induced subprocess.
Currently this has been done only for $gg \to \gamma \gamma$~\cite{Bern:2001df,Bern:2002jx},
and the NLO corrections have been found to be not only sizeable but also
important for stabilising the theoretical predictions~\cite{Bern:2002jx}.
Finally, precise theoretical predictions for $gg \to ZZ$ can be useful
for constraining the total Higgs boson decay 
width at the LHC~\cite{Campbell:2011cu,Kauer:2012hd,Caola:2013yja,Campbell:2013una}.

Technically, the computation of the NLO corrections to $gg \to V_1 V_2$ requires two ingredients, the two-loop
virtual corrections to $gg \to V_1 V_2$ and the one-loop real-virtual corrections to the corresponding
radiative processes with one more parton in the final state.
By now the computation of the one-loop amplitudes with an extra gluon does not constitute any conceptual
difficulty and can be pursued with standard techniques for one-loop 
multi-legs processes~\cite{Berger:2008sj,Hirschi:2011pa,Ellis:2011cr,Bevilacqua:2011xh,Cullen:2011ac,Cascioli:2011va,Cullen:2014yla}.
The two-loop amplitudes, on the other hand, are known only for
$gg \to \gamma \gamma$~\cite{Bern:2001df} and for $gg \to Z \gamma$~\cite{Gehrmann:2013vga},
in both cases for on-shell final state photons.
In order to obtain physical predictions, both contributions need to be combined
using a subtraction scheme to isolate and cancel unphysical
IR divergences. In this case, a NLO scheme~\cite{Frixione:1995ms,Catani:1996vz} would be sufficient. 

In this paper we calculate the missing two-loop massless QCD corrections to
 $gg \to V_1 V_2 $, with an off-shell vector boson pair $V_1 V_2 = \gamma^* \gamma^*, ZZ, Z\gamma^*, W^+W^-$. 
The calculation builds upon the master integrals for four-point 
functions with massless propagators and two massive external legs, which were computed recently 
in the case of equal masses in~\cite{Gehrmann:2013cxs,Gehrmann:2014bfa}, and in
the case of different masses 
in~\cite{Henn:2014lfa,Caola:2014lpa,Papadopoulos:2014hla,Gehrmann:2015ora}.
The former  
were used for the first NNLO fully-inclusive calculations of $ZZ$~\cite{Cascioli:2014yka}
and $W^+W^-$~\cite{Gehrmann:2014fva} production 
at the LHC, while the latter allowed
the computation of the two-loop corrections to
$q \bar{q}' \to V_1 V_2$~\cite{Caola:2014iua,Gehrmann:2015ora}. A subset of these
master integrals was also computed independently in~\cite{Chavez:2012kn,Anastasiou:2014nha}.
While the inclusion of massive top-loop mediated subprocesses would be of interest for
some phenomenological applications~\cite{Campbell:2011cu,Melnikov:2015laa}, the
computation of the two-loop amplitudes requires knowledge of challenging new
master integrals, which should be addressed in the future.

The paper is structured as follows. In Section~\ref{sec:tens} we describe
the tensor decomposition of the partonic current for the process $gg \to V_1 V_2$
and consider the possible electroweak coupling structures.
We include the vector boson decays and describe
the helicity amplitudes for the process $gg \to V_1 V_2\to 4~\mathrm{leptons}$
in terms of scalar form factors in Section~\ref{sec:helamp}.
The actual calculation of the loop contributions to these form factors is
described in Section~\ref{sec:calc}, which includes a dicussion of UV renormalisation,
IR subtraction and various checks we performed on our results.
In Section~\ref{sec:num} we present numerical results obtained with our {\tt C++} implementation.
Finally, we conclude in Section~\ref{sec:conc}.
In Appendix~\ref{sec:form} we give explicit formulae for
obtaining the physical form factors appearing in the helicity amplitudes from the
original tensor coefficients computed in this paper.
We provide computer readable files for our analytical results
and our {\tt C++} code for the numerical evaluation of the amplitudes
on our {\tt VVamp} project page on \hepforge\ at \url{http://vvamp.hepforge.org}.

\section{\texorpdfstring
{Partonic current for $gg\to V_1 V_2$}
{Partonic current for gg -> V1 V2}
}
\label{sec:tens}

We consider the production of two massive off-shell vector bosons, $V_1V_2$, in the gluon fusion channel,
\begin{equation}\label{VVprocess}
 g(p_1) + g(p_2) \longrightarrow V_1(p_3) + V_2(p_4),
\end{equation}
where $V_1V_2$ = $\gamma^* \gamma^*$,
$ZZ$, $Z\gamma^*$,  $W^+W^-$. 
The final states $W^\pm \gamma^*$ and $W^\pm Z$ instead are forbidden by charge conservation.
Since the two vector bosons are off-shell we have in the general case
\begin{equation}
 p_1^2=p_2^2 =0\,, \quad p_3^2>0,\quad p_4^2>0,\quad p_3^2 \neq p_4^2,
\end{equation}
with the usual Mandelstam invariants defined as
\begin{equation}
s=(p_1+p_2)^2\,, \qquad t=(p_1-p_3)^2\,,\qquad u=(p_2-p_3)^2\,,
\end{equation}
and the relation
\begin{equation}
s+t+u = p_3^2 + p_4^2\,.
\end{equation}
The physical region for the scattering kinematics has the boundary $t\,u = p_3^2\,p_4^2$
and fulfils
\begin{equation}
s \geq \Big(\sqrt{p_3^2} + \sqrt{p_4^2}\Big)^2,\qquad
\frac{1}{2}\big(p_3^2+p_4^2-s -\kappa\big) \leq t \leq  \frac{1}{2}\big(p_3^2+p_4^2-s +\kappa\big)
\end{equation}
where $\kappa$ is the K\"all\'en function
\begin{equation}\label{kaellen}
\kappa\left(s,p_3^2,p_4^2\right) \equiv
  \sqrt{s^2 + p_3^4 + p_4^4 - 2 (s\,p_3^2 + p_3^2\,p_4^2 + p_4^2\,s)}\,.
\end{equation}
We denote the scattering amplitude for the process~\eqref{VVprocess} by
\[
S(p_1,p_2,p_3) = S_{\mu\nu\rho\sigma}(p_1,p_2,p_3)\,
 \epsilon_1^{\rho}(p_1)\,\epsilon_2^{\sigma}(p_2)\,
 \epsilon_3^{\ast\,\mu}(p_3)\,\epsilon_4^{\ast\,\nu}(p_4)
\]
where $\epsilon_1$, $\epsilon_2$ are the polarisation vectors of the incoming gluons,
$\epsilon_3$, $\epsilon_4$ are the polarisation vectors of the outgoing massive vector bosons,
$p_4=p_1+p_2-p_3$ and an overall factor $e^2$ is kept implicit
with $e$ being the positron charge.
Since we will consider leptonic decays of the massive vector bosons we will be
able to construct the full amplitude including the decays from
the partonic current
\[
S_{\mu\nu}(p_1,p_2,p_3) = S_{\mu \nu\rho\sigma}(p_1,p_2,p_3)
 \,\epsilon_1^\rho(p_1) \,\epsilon_2^\sigma(p_2)
\]
for the $2\to 2$ process.
In particular, it is only the latter which receives (pure) QCD corrections
at any order in perturbation theory.

In order to compute the partonic current
it is useful to consider its tensor decomposition.
Based on Lorentz invariance only, there are 138 independent tensor structures which can contribute
\begin{align}
S^{\mu \nu\rho\sigma}(p_1,p_2,p_3) &=
        a_1 g^{\mu\nu} g^{\rho\sigma}
       +a_2 g^{\mu\rho} g^{\nu\sigma}  
       +a_3 g^{\mu\sigma} g^{\nu\rho} \nonumber \\
&+\sum\limits_{j_1,j_2=1}^{3} 
\Bigl(
   b^{(1)}_{j_1 j_2}\,g^{\mu\nu}\,p_{j_1}^{\rho}\,p_{j_2}^{\sigma}
  +b^{(2)}_{j_1 j_2}\,g^{\mu\rho}\,p_{j_1}^{\nu}\,p_{j_2}^{\sigma}
  +b^{(3)}_{j_1 j_2}\,g^{\mu\sigma}\,p_{j_1}^{\nu}\,p_{j_2}^{\rho}\nonumber \\ 
&\phantom{\sum\limits_{j_1,j_2=1}^{3}}
+b^{(4)}_{j_1 j_2}\,g^{\nu\rho}\,p_{j_1}^{\mu}\,p_{j_2}^{\sigma}
  +b^{(5)}_{j_1 j_2}\,g^{\nu\sigma}\,p_{j_1}^{\mu}\,p_{j_2}^{\rho} 
  +b^{(6)}_{j_1 j_2}\,g^{\rho\sigma}\,p_{j_1}^{\mu}\,p_{j_2}^{\nu}
\Bigr) 
\nonumber \\  
 &+\sum\limits_{j_1,j_2,j_3,j_4=1}^{3} c_{j_1 j_2 j_3 j_4}
p_{j_1}^{\mu}\,p_{j_2}^{\nu} p_{j_3}^{\rho}\,p_{j_4}^{\sigma}, \label{tensordec1}
\end{align}
where the coefficients $a_j$, $b_{ij}^k$ and $c_{ijkl}$ are scalar functions of the
kinematic invariants $s$, $t$, $p_3^2$, $p_4^2$ and of the space-time dimension $d$.
Not all structures are relevant for our calculation. 
Many of them simply drop
due to the transversality of the gluons' polarisation vectors
\begin{equation}
\epsilon_1 \cdot p_1 = \epsilon_2 \cdot p_2 = 0\,.\label{transverseg}
\end{equation}
Moreover the tensor structure can be further simplified by fixing explicitly
the gauge for the incoming gluons. A particularly simple choice is given by the symmetrical
condition
\begin{align}
 \epsilon_1 \cdot p_2 &= \epsilon_2 \cdot p_1 = 0\,, \label{gaugeg}
\end{align}
which corresponds to the following rules for the polarisation sums
\begin{align}
 \sum_{\lambda_1} \epsilon_{1\,\lambda_1}^{\mu *}(p_1) \epsilon_{1\,\lambda_1}^{\nu}(p_1)
 &= -g^{\mu \nu} + \frac{p_1^\mu p_2^\nu + p_1^\nu p_2^\mu }{p_1 \cdot p_2}\,,\nonumber \\
  \sum_{\lambda_2} \epsilon_{2\,\lambda_2}^{\mu *}(p_2) \epsilon_{2\,\lambda_2}^{\nu}(p_2)
 &= -g^{\mu \nu} + \frac{p_1^\mu p_2^\nu + p_1^\nu p_2^\mu }{p_1 \cdot p_2}\,. \label{polsum12}
\end{align}
Further conditions can be applied on the polarisation vectors of the massive vector bosons $V_1$, $V_2$.
We employ for their polarisation vectors
\begin{align}
 \epsilon_3 \cdot p_3 = \epsilon_4 \cdot p_4 = 0\,, \label{gaugeV}
\end{align}
and for the polarisation sums
\begin{align}
 \sum_{\lambda_3} \epsilon_{3\,\lambda_3}^{\mu *}(p_3) \epsilon_{3\,\lambda_3}^{\nu}(p_3)
 &= -g^{\mu \nu} + \frac{p_3^\mu p_3^\nu }{p_3^2 }\,,\nonumber \\
  \sum_{\lambda_4} \epsilon_{4\,\lambda_4}^{\mu *}(p_4) \epsilon_{4\,\lambda_4}^{\nu}(p_4)
 &= -g^{\mu \nu} + \frac{p_4^\mu p_4^\nu }{p_4^2 }\,. \label{polsum34}
\end{align}

Imposing the constraints~\eqref{transverseg}, \eqref{gaugeg} and \eqref{gaugeV}
one is left with only $20$ independent tensor structures and
we can write the partonic current according to
\begin{align} 
S^{\mu \nu}(p_1,p_2,p_3)
&= 
\sum_{j=1}^{20}\, A_j(s,t,p_3^2,p_4^2)\, T_j^{\mu \nu}\,,\label{tensordec}
\end{align}
where the $A_j$ are scalar functions of $s$, $t$, $p_3^2$, $p_4^2$ and $d$.
The tensors $T^{\mu \nu}_j$ are defined as
\begin{gather}\label{tensors}
\begin{aligned}
 T_1^{\mu \nu} &= \epsilon_1 \cdot \epsilon_2\, g^{\mu \nu}\,, &
 T_2^{\mu \nu}  &= \epsilon_1^{\mu}\, \epsilon_2^{\nu}\,, &
 T_3^{\mu \nu}  &= \epsilon_1^{\nu}\, \epsilon_2^{\mu}\,, &
 T_4^{\mu \nu} &= \epsilon_1 \cdot \epsilon_2\,p_1^{\mu}\, p_1^{\nu}\,,\\
 T_5^{\mu \nu}  &= \epsilon_1 \cdot \epsilon_2\,p_1^{\mu}\, p_2^{\nu}\,,&
 T_6^{\mu \nu}  &= \epsilon_1 \cdot \epsilon_2\,p_2^{\mu}\, p_1^{\nu}\,,&
 T_7^{\mu \nu}  &= \epsilon_1 \cdot \epsilon_2\,p_2^{\mu}\, p_2^{\nu}\,,&
 T_8^{\mu \nu}   &= \epsilon_2 \cdot p_3\,\epsilon_1^\mu \,p_1^{\nu}\,,\\
 T_9^{\mu \nu}    &= \epsilon_2 \cdot p_3\,\epsilon_1^\mu \,p_2^{\nu}\,,&
 T_{10}^{\mu \nu} &= \epsilon_2 \cdot p_3\,\epsilon_1^\nu \,p_1^{\mu}\,,&
 T_{11}^{\mu \nu} &= \epsilon_2 \cdot p_3\,\epsilon_1^\nu \,p_2^{\mu}\,,&
 T_{12}^{\mu \nu} &= \epsilon_1 \cdot p_3\,\epsilon_2^\mu \,p_1^{\nu}\,,\\
 T_{13}^{\mu \nu}  &= \epsilon_1 \cdot p_3\,\epsilon_2^\mu \,p_2^{\nu}\,,&
 T_{14}^{\mu \nu}  &= \epsilon_1 \cdot p_3\,\epsilon_2^\nu \,p_1^{\mu}\,,&
 T_{15}^{\mu \nu}  &= \epsilon_1 \cdot p_3\,\epsilon_2^\nu \,p_2^{\mu}\,,&
 T_{16}^{\mu \nu} &= \epsilon_1 \cdot p_3\,\epsilon_2 \cdot p_3\, g^{\mu \nu}\,,
\end{aligned}
\nonumber\\
\begin{aligned}
 T_{17}^{\mu \nu} &= \epsilon_1 \cdot p_3\,\epsilon_2 \cdot p_3\,p_1^\mu \,p_1^{\nu}\,, &
 T_{18}^{\mu \nu}  &= \epsilon_1 \cdot p_3\,\epsilon_2 \cdot p_3\,p_1^\mu \,p_2^{\nu}\,,
\end{aligned}
\nonumber\\
\begin{aligned}
 T_{19}^{\mu \nu}  &= \epsilon_1 \cdot p_3\,\epsilon_2 \cdot p_3\,p_2^\mu \,p_1^{\nu}\,, &
 T_{20}^{\mu \nu}  &= \epsilon_1 \cdot p_3\,\epsilon_2 \cdot p_3\,p_2^\mu \,p_2^{\nu}\,.
\end{aligned}
\end{gather}
We stress that the tensor decomposition~\eqref{tensordec} is based only 
on Lorentz symmetry, gauge invariance and the properties of the boson decays
and holds therefore at every order in perturbative QCD.
Moreover, no assumption has been made on the dimensionality of space-time and the
result is valid for any values of the parameter $d$.

The scalar form factors $A_j$ can be extracted from the amplitude~\eqref{tensordec}
by applying suitable projecting operators. The projectors themselves can be decomposed in the same
$20$ tensors as
\begin{align}
 P_j^{\mu \nu} = \sum_{i=1}^{20} B_{ji} \left( T_i^{\mu \nu}\right)^\dagger \qquad \text{for~}j=1,\ldots,20, \label{proj}
\end{align}
where also $B_{ji}$ are functions of the external invariants and $d$.
Their explicit form can be determined imposing
\begin{align}
 \sum_{pol} P_j^{\mu' \nu'} \left[ \epsilon_{3 \mu'} \epsilon_{4 \nu'}
 \epsilon_{3 \mu}^* \epsilon_{4 \nu}^* \right]
 S^{\mu \nu} 
 &= A_j \quad \text{for~}j=1,...,20,
\end{align}
where the polarisation sums are evaluated in $d$ dimensions according to \eqref{polsum12} and \eqref{polsum34}.
The explicit results for the coefficients $B_{ji}$ are rather lengthy and we prefer not to write them here
explicitly. Computer readable files for the latter are given on our project page at \hepforge.

\begin{figure}
\centerline{\includegraphics[width=0.95\linewidth]{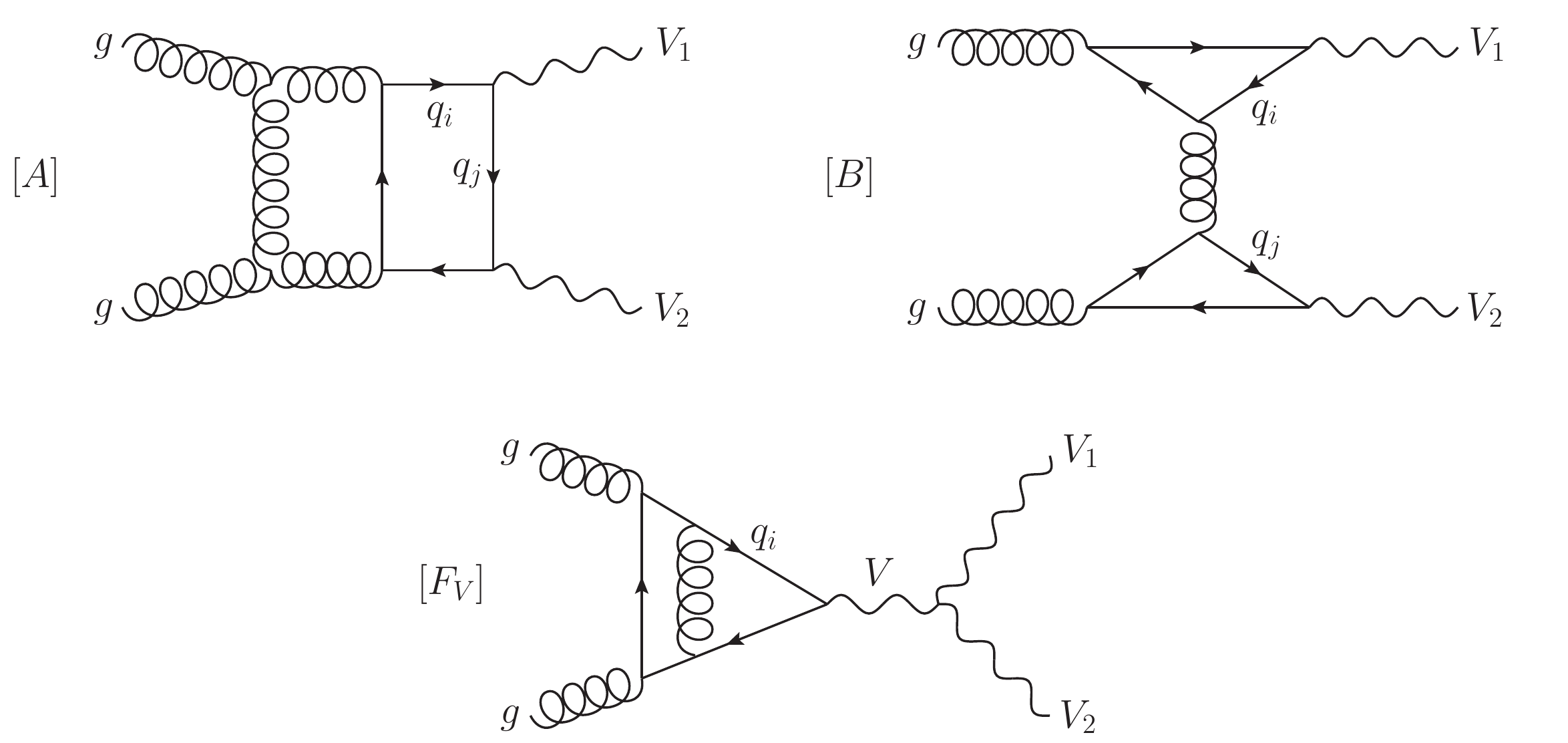}}
\caption{\label{fig:diagclass}
Example Feynman diagrams for the process $gg \to V_1 V_2$ at the two-loop
level, where the vector bosons couple to the same fermion loop, $[A]$, to
different fermion loops, $[B]$, or to an intermediate vector boson, $[F_V]$.
The sum of all type $[B]$ contribution and the sum of all type $[F_V]$
contributions vanish, respectively.
}
\end{figure}

The partonic current is the only one which receives contributions from
QCD radiative corrections and, for two gluons of helicities $\lambda_1$ and $\lambda_2$,
can be written as
\begin{equation}
 S_{\mu \nu}(p_1^{\lambda_1},p_2^{\lambda_2},p_3) = \delta^{a_1 a_2}\,\sum_{j}
 \C_{V_1 V_2}^{[j]}\,S^{[j]}_{\mu \nu \rho \sigma}(p_1,p_2,p_3) 
 \epsilon_{1 \lambda_1 }^{\rho}(p_1)\epsilon_{2 \lambda_2}^{\sigma }(p_2)\,,
 \label{ParCurr1}
\end{equation}
where
$\delta^{a_1 a_2}$ is the overall colour structure and
the index $j$ runs over different possible classes of diagrams 
discussed below, see also Fig.~\ref{fig:diagclass}, which are characterised by
different electroweak couplings $\C^{[j]}_{V_1 V_2}$. 

Before proceeding, it is convenient to introduce some notations needed in the
following.
As long as we work in QCD, we only need to consider
the coupling of electroweak vector bosons $V$ to fermions.
We follow~\cite{Denner:1991kt} and parametrise the couplings as
\begin{equation}
\mathcal{V}_\mu^{V f_1 f_2} = i\, e\, \Gamma_\mu^{V f_1 f_2}\,, \qquad 
\mbox{where} \quad e = \sqrt{4\,\pi\,\alpha} \quad \mbox{is the positron charge}\,, 
\end{equation}
such that all fermion charges are expressed in units of $e$ and 
\begin{equation}
\Gamma_\mu^{V f_1 f_2} = L_{f_1 f_2}^V \, \gamma_\mu \left( \frac{1-\gamma_5}{2}\right)
+ R_{f_1 f_2}^V \, \gamma_\mu \left( \frac{1+\gamma_5}{2}\right)\,,
\end{equation}
with
\begin{alignat}{2}
L_{f_1f_2}^\gamma &= -e_{f_1} \,\delta_{f_1f_2} &
R_{f_1f_2}^\gamma &= -e_{f_1} \,\delta_{f_1f_2}\,,
\label{gLBcoupl}
\\
L_{f_1f_2}^Z &= \frac{I_3^{f_1} - \sin^2 {\theta_w} e_{f_1}}{\sin{\theta_w} \cos{\theta_w}} \,\delta_{f_1f_2}\,, &\qquad
R_{f_1f_2}^Z &= -\frac{\sin{\theta_w} e_{f_1}}{\cos{\theta_w}} \,\delta_{f_1f_2}\,,
\label{ZLRcoupl}
\\
L_{f_1f_2}^W &= \frac{1 }{\sqrt{2}\, \sin{\theta_w}} \,\epsilon_{f_1f_2} \,,&
R_{f_1f_2}^W &= 0\,, \label{WLRcoupl}
\end{alignat}
where $\epsilon_{f_1f_2}$ is unity for $f_1\neq f_2$, but belonging to the same isospin doublet, 
and zero otherwise. 

Let us consider the different electroweak coupling structures in detail.
It is clear that, since we do not take any electroweak radiative corrections into account, 
at least one of the two vector bosons must be 
coupled to an internal fermion loop.
In order to compute the one- and two-loop massless QCD corrections
we need to consider the following three possibilities, see Fig.~\ref{fig:diagclass}.
\begin{description}
 \item[Class $\mathbf{A}$:] Both vector bosons $V_1 V_2$ are attached to the same fermion loop.
 In this case the diagrams 
 are proportional to the charge weighted sum of the quark flavours, which we denote as
$\C_{V_1 V_2}^{[A]} = N_{V_1 V_2}$. 
These diagrams could in principle yield two different contributions.
One, proportional to the sum of the vector-vector and the axial-axial couplings,
in which all dependence on $\gamma_5$ cancels out. The second, instead, contains
the vector-axial coupling and is linear in $\gamma_5$. 
Due to charge parity conservation this last contribution is expected to always vanish 
identically for massless 
quarks running in the loops, 
for any choice of $V_1$ and $V_2$~\cite{Glover:1988rg,Glover:1988fe,Melnikov:2015laa}.
One then easily finds that
\begin{align}\label{coupl1}
N_{\gamma \gamma} &= \frac{1}{2}\sum_{i} \left[ \left( L_{q_i q_i}^\gamma \right)^2 + \left( R_{q_i q_i}^\gamma \right)^2 \right], &
N_{Z \gamma} &=  \frac{1}{2} \sum_{i} \left( L_{q_i q_i}^Z L_{q_i q_i}^\gamma + R_{q_i q_i}^Z R_{q_i q_i}^\gamma \right), \nonumber\\
N_{ZZ} &= \frac{1}{2} \sum_{i} \left[  \left( L_{q_i q_i}^Z\right)^2 + \left(R_{q_i q_i}^Z \right)^2 \right], &
N_{WW} &= \frac{1}{2} \sum_{i,\,j} \left( L_{q_i q_j}^W  L_{q_j q_i}^W \right),
\end{align}
where the indices $i,j$ run over the flavours of the quarks in the loop and
$L_{q_i q_i}^\gamma = R_{q_i q_i}^\gamma$ such that $N_{\gamma\gamma}=\sum_{i} e_{q_i}^2$.

\item[Class $\mathbf{B}$:] The two vector bosons are attached to two different fermion loops.
This configuration is of course possible only starting from two loops on.
Each fermion loop contains both a vector and an axial piece.
For the case of two-loop massless QCD corrections relevant here,
both contributions can be shown to vanish.
The axial contribution cancels out for degenerate isospin doublets, while the vector
piece must sum up to zero due to Furry's theorem.

\item[Classes $\mathbf{F_V}$:] Only for the case of $V_1 V_2 = W^+W^-$, one should also take into
account the $s$-channel production diagrams, where the incoming gluons produce an intermediate electroweak
gauge boson $V=\gamma^*/Z^*$, which then decays into the outgoing $W$-pair, see Fig.~\ref{fig:diagclass}.
Charge-parity invariance ensures that the vector part of these diagrams must sum up to zero.
Again, the axial part cancels out for degenerate isospin doublets, and therefore also in
the case of massless quarks running in the loops.
\end{description}
For the case of the one- and two-loop contributions considered here, we can therefore
simplify~\eqref{ParCurr1} to
\begin{align}
 S_{\mu \nu}(p_1^{\lambda_1},p_2^{\lambda_2},p_3) &= 
 \delta^{a_1 a_2}\,N_{V_1 V_2}\,S^{[A]}_{\mu \nu \rho \sigma}(p_1,p_2,p_3) 
  \epsilon_{1 \lambda_1 }^{\rho}(p_1)\epsilon_{2 \lambda_2}^{\sigma }(p_2)
\,,
 \label{ParCurr2}
\end{align}
with $N_{V_1 V_2}$ given in~\eqref{coupl1}
and consider the coefficients $A_j^{[A]}$ defined by
\begin{align}
A_j(s,t,p_3^2,p_4^2) &= \delta_{a_1 a_2} N_{V_1V_2} A_j^{[A]}(s,t,p_3^2,p_4^2).
\end{align}

It is instructive to study the transformations of the partonic current~\eqref{ParCurr2} 
under permutations of the external legs. We define the following two permutations
\begin{align}
&\pi_{12} := p_1 \leftrightarrow p_2 \Rightarrow \{\,t \leftrightarrow u \,\}, \nonumber \\
&\pi_{34} := p_3 \leftrightarrow p_4 \Rightarrow \{\, t \leftrightarrow u\,, \;\; p_3^2 \leftrightarrow p_4^2 \,\} \,.
\label{perm}
\end{align}
Because of Bose symmetry these two permutations must leave the partonic amplitude unchanged.
This enforces a well defined behaviour of the coefficients $A_j(s,t,p_3^2,p_4^2)$ under the action
of $\pi_{12}$ and $\pi_{34}$. From direct inspection of~\eqref{tensordec}  one finds that the following
relations must be fulfilled:
\begin{alignat}{3}
\pi_{12}:\quad
A^{[A]}_1(s,u,p_3^2,p_4^2) &= A^{[A]}_1(s,t,p_3^2,p_4^2)\,,
\qquad  &
A^{[A]}_2(s,u,p_3^2,p_4^2) &= A^{[A]}_3(s,t,p_3^2,p_4^2)\,, 
\nonumber \\
A^{[A]}_4(s,u,p_3^2,p_4^2) &= A^{[A]}_7(s,t,p_3^2,p_4^2)\,, 
&
A^{[A]}_5(s,u,p_3^2,p_4^2) &= A^{[A]}_6(s,t,p_3^2,p_4^2)\,,
\nonumber \\
A^{[A]}_8(s,u,p_3^2,p_4^2) &= A^{[A]}_{13}(s,t,p_3^2,p_4^2)\,, 
&
A^{[A]}_{9}(s,u,p_3^2,p_4^2) &= A^{[A]}_{12}(s,t,p_3^2,p_4^2)\,,
\nonumber \\
A^{[A]}_{10}(s,u,p_3^2,p_4^2) &= A^{[A]}_{15}(s,t,p_3^2,p_4^2)\,,
&
A^{[A]}_{11}(s,u,p_3^2,p_4^2) &= A^{[A]}_{14}(s,t,p_3^2,p_4^2)\,,
\nonumber \\
A^{[A]}_{16}(s,u,p_3^2,p_4^2) &= A^{[A]}_{16}(s,t,p_3^2,p_4^2)\,,
&
A^{[A]}_{17}(s,u,p_3^2,p_4^2) &= A^{[A]}_{20}(s,t,p_3^2,p_4^2)\,,
\nonumber \\
A^{[A]}_{18}(s,u,p_3^2,p_4^2) &= A^{[A]}_{19}(s,t,p_3^2,p_4^2)\,,
&&\label{Ajx12}
\\[1ex]
\pi_{34}:\quad
A^{[A]}_1(s,u,p_4^2,p_3^2) &= A^{[A]}_1(s,t,p_3^2,p_4^2)\,,
&
A^{[A]}_2(s,u,p_4^2,p_3^2) &= A^{[A]}_3(s,t,p_3^2,p_4^2)\,,
\nonumber \\
A^{[A]}_4(s,u,p_4^2,p_3^2) &= A^{[A]}_4(s,t,p_3^2,p_4^2)\,, 
&
A^{[A]}_5(s,u,p_4^2,p_3^2) &= A^{[A]}_6(s,t,p_3^2,p_4^2)\,,
\nonumber \\
A^{[A]}_7(s,u,p_4^2,p_3^2) &= A^{[A]}_7(s,t,p_3^2,p_4^2)\,, 
&
A^{[A]}_{8}(s,u,p_4^2,p_3^2) &= - A^{[A]}_{10}(s,t,p_3^2,p_4^2)\,,
\nonumber \\
A^{[A]}_{9}(s,u,p_4^2,p_3^2) &= - A^{[A]}_{11}(s,t,p_3^2,p_4^2)\,,
&
A^{[A]}_{12}(s,u,p_4^2,p_3^2) &= - A^{[A]}_{14}(s,t,p_3^2,p_4^2)\,,
\nonumber \\
A^{[A]}_{13}(s,u,p_4^2,p_3^2) &= - A^{[A]}_{15}(s,t,p_3^2,p_4^2)\,,
&
A^{[A]}_{16}(s,u,p_4^2,p_3^2) &= A^{[A]}_{16}(s,t,p_3^2,p_4^2)\,,
\nonumber \\
A^{[A]}_{17}(s,u,p_4^2,p_3^2) &= A^{[A]}_{17}(s,t,p_3^2,p_4^2)\,, 
&
A^{[A]}_{18}(s,u,p_4^2,p_3^2) &= A^{[A]}_{19}(s,t,p_3^2,p_4^2)\,,
\nonumber \\
A^{[A]}_{20}(s,u,p_4^2,p_3^2) &= A^{[A]}_{20}(s,t,p_3^2,p_4^2)\,,
&& \label{Ajx34}
\end{alignat}
It is interesting to notice that, upon exploiting all of these crossing relations,
only $9$ out of the $20$ coefficients $A_j^{[A]}$ turn out to be effectively
independent, while the other $11$ coefficients can be obtained by crossing of
the external legs.

\section{\texorpdfstring
{Helicity amplitudes for $gg \to V_1 V_2 \to 4~\mathrm{leptons}$}
{Helicity amplitudes for gg -> V1 V2 -> 4 leptons}
}
\label{sec:helamp}

We consider physical processes, where the two off-shell vector bosons
decay into lepton pairs
\begin{equation}
 g(p_1) + g(p_2) \rightarrow V_1(p_3) + V_2(p_4) \rightarrow l_5(p_5) + \bar{l}_6(p_6) + l_7(p_7) + \bar{l}_8(p_8)
 \label{ggVVllll}
\end{equation}
such that
$p_3= p_5+p_6\,$, $\,p_4=p_7+p_8$ and $p_5^2=p_6^2=p_7^2=p_8^2 = 0$.
%
As long as we consider QCD radiative corrections
the amplitudes $\mathcal{M}_{\lambda_1 \lambda_2 \lambda_3 \lambda_4}^{V_1 V_2}$ can be written, at any
order in perturbation theory, as the product of the partonic current for
$gg \to V_1 V_2$ with the two leptonic currents for the decay products,
$V_1 \to l_5 \bar{l}_6$ and $V_2 \to l_7 \bar{l}_8$,
mediated by the propagators of the two off-shell vector bosons $V_1$ and $V_2$.
We write the propagator for an off-shell vector boson in the $R_\xi$ gauge as
\begin{equation}
P_{\mu \nu}^V(q) = \frac{i\, \Delta^V_{\mu \nu}(q,\xi)}{D_V(q)}\,, \label{prop}
\end{equation}
with 
\begin{equation}
\Delta^V_{\mu \nu}(q,\xi) = \left( - g_{\mu \nu} + (1 - \xi) \frac{q_\mu q_\nu}{q^2 - \xi m_V^2}\right)\,,
\label{numprop}
\end{equation}
\begin{align}
&D_{\gamma^*}(q) = q^2\,,\qquad D_{Z,W}(q) = (q^2 - m_V^2 + i\,\Gamma_V m_V)\,,\label{denprop}
\end{align}
where $m_V$ is its mass and $\Gamma_V$ is its decay width.
In our case the massive vector bosons couple to massless fermion lines such that the
term proportional to $(1-\xi)$ can be dropped.

In the following we consider fixed helicities of the external particles
and compute the amplitudes for the different helicity configurations.
While the quantities and formulae presented up to this point were treated in $d$ dimensions
throughout, we now consider 4-dimensional external states in order to compute
the amplitudes for specific helicities.
Since the decay leptons are massless, helicity is conserved along the leptonic decay currents
and the amplitude can be written as
\begin{equation}
\mathcal{M}_{\lambda_1 \lambda_2 \lambda_3 \lambda_4}^{V_1 V_2}(p_1, p_2;p_5,p_6,p_7,p_8),
\end{equation}
where $\lambda_1$ and $\lambda_2$ are the helicities of the incoming gluons, while
$\lambda_3$ and $\lambda_4$ are the helicities of the two 
leptonic currents. It is clear that there are 16 different helicity configurations,
depending on the different possibilities for the initial and final states.
Each gluon has two possible helicity states, which we denote
by $L$ (-) and $R$ (+), and similarly each leptonic current occurs in either left- or
right-handed configuration, again denoted by $L$ and $R$, respectively, such that
$\lambda_j = L,R$, for $j=1,...,4$. As we will show explicitly later on,
all 16 helicity configurations can be obtained from only two independent ones, by simple 
permutations of the external legs and complex conjugation.
We choose as independent configurations the following two
\begin{equation}
\mathcal{M}_{LLLL}^{V_1 V_2}(p_1, p_2;p_5,p_6,p_7,p_8)\,, \qquad 
\mathcal{M}_{LRLL}^{V_1 V_2}(p_1, p_2;p_5,p_6,p_7,p_8)\,. \label{helconf}
\end{equation}

With the notations introduced above we write the two independent helicity amplitudes~\eqref{helconf},
up to two loops, as:
\begin{align}
 \mathcal{M}_{\lambda_1 \lambda_2 LL}^{V_1 V_2}(p_1, p_2;p_5,p_6,p_7,p_8) &= 
 (4 \pi \alpha)^2\,
 \frac{L_{f_5f_6}^{V_1}\,L_{f_7f_8}^{V_2}}{D_{V_1}(p_3)D_{V_2}(p_4)}\,
 \M_{\lambda_1 \lambda_2 LL}(p_1, p_2;p_5,p_6,p_7,p_8)\,,\label{helampl}
\end{align}
where
the \textsl{basic amplitudes} $\M_{\lambda_1 \lambda_2 LL}(p_1, p_2;p_5,p_6,p_7,p_8)$
are constructed from the partonic current~\eqref{ParCurr1} and
the leptonic currents~\eqref{LepCurrL} according to
\begin{align}
 \M_{\lambda_1 \lambda_2 LL}(p_1, p_2;p_5,p_6,p_7,p_8) = 
 \epsilon_{1 \lambda_1 }^{\rho}(p_1)\epsilon_{2 \lambda_2}^{\sigma }(p_2)
 \,S_{\mu \nu\rho \sigma}(p_1,p_2,p_3)\,L_{L}^{\mu}(p_5^-,p_6^+)\,L_{L}^{\nu}(p_7^-,p_8^+)\,. \label{Bhelampl}
\end{align}
The leptonic decay currents do not receive
any QCD corrections and are simple tree-level objects. They can be easily
expressed in the usual spinor-helicity notation~\cite{Dixon:1996wi,Dixon:1998py} as
\begin{align}
&L^\mu_L(p_5^-,p_6^+) = \bar{u}_-(p_5) \,\gamma^{\mu} \, v_+(p_6) = [ 6\, | \gamma^{\mu } | \,5\, \rangle 
= \langle 5\, | \gamma^{\mu } | \,6\, ]  \,,\label{LepCurrL}\\
&L^\mu_R(p_5^+,p_6^-) = \bar{u}_+(p_5) \,\gamma^{\mu}\, v_-(p_6)=  [ 5\, | \gamma^{\mu } | \,6\, \rangle 
= \left( L^\mu_L(p_5^-,p_6^+) \right)^* = L_L^\mu(p_6^-,p_5^+)\,. \label{LepCurrR}
\end{align}
Note that, in this case, a permutation of the external
momenta is equivalent to a complex conjugation of the current and
it corresponds to a flip of the helicity $L \leftrightarrow R$.

Once the tensor decomposition of the partonic current is fixed, it is straight-forward
to express the two basic helicity amplitudes $\M_{LLLL}$ and $\M_{LRLL}$ in~\eqref{Bhelampl}
in the usual spinor-helicity notation~\cite{Dixon:1996wi,Dixon:1998py}.
We replace the gluon polarisation vectors according to
\begin{align}
 \epsilon_{1L}^\mu(p_1) = \frac{[ 2 | \gamma^\mu | 1 \rangle }{\sqrt{2} [12]}\,, \quad
 \epsilon_{1R}^\mu(p_1) =  \frac{\langle 2 | \gamma^\mu | 1 ] }{\sqrt{2} \langle 21 \rangle }\,, \quad
 \epsilon_{2L}^\mu(p_2) = \frac{[ 1 | \gamma^\mu | 2 \rangle }{\sqrt{2} [21]}\,, \quad
 \epsilon_{2R}^\mu(p_2) =  \frac{\langle 1 | \gamma^\mu | 2 ] }{\sqrt{2} \langle 12 \rangle }\,,
 \label{polvec}
\end{align}
which is of course compatible with the polarisation sums~\eqref{polsum12} and \eqref{polsum34}.
Note again that here we are assuming $4$-dimensional external states. This allows to
reduce considerably the number of independent structures that are required for 
parametrising a specific helicity configuration.
Using~\eqref{polvec} we find that both basic amplitudes can be written in terms of $9$ independent
spinor structures as
\begin{align}
\M_{\lambda_1 \lambda_2 LL}(p_1,p_2;p_5,p_6,p_7,p_8) &= 
 C_{\lambda_1 \lambda_2} \Bigg\{ [ 2\, \p_3\, 1 \rangle\, \Big(
   E_1^{\lambda_1 \lambda_2} \langle 5 7 \rangle [ 68 ] \nonumber \\
 &+ E_2^{\lambda_1 \lambda_2}\, \langle 15 \rangle \langle 17 \rangle [16][18] 
+ E_3^{\lambda_1 \lambda_2}\, \langle 15 \rangle \langle 27 \rangle [16][28] \nonumber \\
&+ E_4^{\lambda_1 \lambda_2}\, \langle 25 \rangle \langle 17 \rangle [26][18]
 +
 E_5^{\lambda_1 \lambda_2}\, \langle 25 \rangle \langle 27 \rangle [26][28] \,
  \Big) \nonumber \\
&+ E_6^{\lambda_1 \lambda_2}\, \langle 15 \rangle \langle 17 \rangle [16][28] 
+   E_7^{\lambda_1 \lambda_2}\, \langle 15 \rangle \langle 17 \rangle [26][18] \nonumber \\
&+ E_8^{\lambda_1 \lambda_2}\, \langle 15 \rangle \langle 27 \rangle [26][28] 
+ E_9^{\lambda_1 \lambda_2}\, \langle 25 \rangle \langle 17 \rangle [26][28] \Bigg\} \,, \label{MllLL}
\end{align}
where the $18$ newly introduced form factors $E_j^{\lambda_1 \lambda_2}$
are simple linear combinations of the scalar coefficients $A_j$. 
The spinor structure of the amplitudes for the configurations $LLLL$ and $LRLL$
differs only by an overall factor which reads in the two cases
\begin{align}\label{cll}
 C_{LL} = [1\,\p_3\,2\rangle \frac{\langle 1 2 \rangle }{[12]}\,,\qquad
 C_{LR} = [2\,\p_3\,1\rangle\,,
\end{align}
but the form factors $E_j^{LL}$ and $E_j^{LR}$ are different.
We also note, in passing, that the spinor structure of~\eqref{MllLL} exhibits also
a \emph{formal} similarity to that of the $RLL$ amplitude for
$q \bar{q}' \to V_1 V_2 \to l_5\bar{l}_6 l_7 \bar{l}_8$~\cite{Caola:2014iua,Gehrmann:2015ora},
again up to an overall factor and with, of course, completely unrelated form factors.
Similar as before, we also define the functions $E_j^{\lambda_1\lambda_2\,[A]}$
\begin{equation}
E_j^{\lambda_1\lambda_2}(s,t,p_3^2,p_4^2) = \delta_{a_1 a_2} N_{V_1V_2} E_j^{\lambda_1\lambda_2\,[A]}(s,t,p_3^2,p_4^2).
\end{equation}
The explicit expressions for the form factors $E_j^{\lambda_1\lambda_2}$ in terms
of the coefficients $A_j$ are given in Appendix~\ref{sec:form}.

In order to obtain all $16$ helicity amplitudes from~\eqref{MllLL}, one should recall that
complex conjugation has the effect of reversing the helicity of the external 
gluons, 
\begin{align}
  \left( \epsilon_{1L}^\mu(p_1) \right)^*  = \epsilon_{1R}^\mu(p_1)\,,& \qquad 
  \left( \epsilon_{2L}^\mu(p_2) \right)^*  = \epsilon_{2R}^\mu(p_2)\,,
 \end{align}
and similarly for the leptonic currents, see~\eqref{LepCurrR}~\eqref{LepCurrL}.
We define with the symbol $[...]^C$ a complex-conjugation operation which, when applied on the amplitudes
$\M_{\lambda_1 \lambda_2 LL}$, acts \textsl{only on the spinor structures}, i.e. leaves
invariant the form factors $E_{j}^{\lambda_1 \lambda_2}$. Given the explicit
form of~\eqref{MllLL}, it is easy to see that this corresponds to simply exchanging
angle brackets with squared bracket and vice versa
\begin{equation}
 \left[ \M_{\lambda_1 \lambda_2 LL} \right]^C
 \equiv \M_{\lambda_1 \lambda_2 LL} \left( \langle ij \rangle \leftrightarrow [ij] \right)\,.
\end{equation}
Hence, we can derive the missing helicity amplitudes for
left-handed leptonic currents from the two basic amplitudes as
\begin{align}
 \M_{RLLL}(p_1,p_2;p_5,p_6,p_7,p_8) &= \left[ \M_{LRLL}(p_1,p_2;p_6,p_5,p_8,p_7) \right]^C\,,\nonumber \\ 
 \M_{RRLL}(p_1,p_2;p_5,p_6,p_7,p_8) &= \left[ \M_{LLLL}(p_1,p_2;p_6,p_5,p_8,p_7) \right]^C\,,
\end{align}
where one should note that the lepton and anti-lepton momenta are exchanged in the r.h.s.
in order to have a left-handed leptonic currents on the l.h.s.
The corresponding formulae for the basic amplitudes for right-handed leptonic currents 
can be obtained from the ones above by simple permutations of the lepton and anti-lepton momenta
\begin{align}
 \M_{\lambda_1 \lambda_2 RL}(p_1,p_2;p_5,p_6,p_7,p_8) &= 
 \M_{\lambda_1 \lambda_2LL}(p_1,p_2;p_6,p_5,p_7,p_8)  \,,\nonumber \\ 
 \M_{\lambda_1 \lambda_2LR}(p_1,p_2;p_5,p_6,p_7,p_8) &= 
 \M_{\lambda_1 \lambda_2LL}(p_1,p_2;p_5,p_6,p_8,p_7)  \,,\nonumber \\ 
 \M_{\lambda_1 \lambda_2RR} (p_1,p_2;p_5,p_6,p_7,p_8)  &= 
\M_{\lambda_1 \lambda_2LL}(p_1,p_2;p_6,p_5,p_8,p_7)  \,. \label{allBhelamp}
\end{align}
With these formulae also all the 16 physical amplitudes 
$\mathcal{M}^{V_1 V_2}_{\lambda_1 \lambda_2 \lambda_3 \lambda_4}$ in~\eqref{helampl}
can be easily obtained, recalling that in the case of right-handed leptonic 
currents one should, of course, exchange the corresponding
couplings $L_{f_i f_j}^{V} \leftrightarrow R_{f_i f_j}^V$.

As we already stated above, the partonic current receives contributions
form QCD radiative corrections and it can be expanded as
\begin{align}
 S_{\mu \nu \rho\sigma}(p_1,p_2,p_3) = 
 \left(\frac{\alpha_s}{2 \pi} \right) S_{\mu \nu \rho\sigma}^{(1)}(p_1,p_2,p_3) 
 + \left(\frac{\alpha_s}{2 \pi} \right)^2 S_{\mu \nu \rho\sigma}^{(2)}(p_1,p_2,p_3)
 + \mathcal{O}(\alpha_s^3)\,,
\end{align}
where obviously the perturbative expansion starts only at one-loop order.
Of course also the coefficients $A_j$, and equivalently the form factors
$E_j^{\lambda_1 \lambda_2}$, have the same expansion
\begin{align}
 A_j &=  \left(\frac{\alpha_s}{2 \pi} \right) A_j^{(1)} 
 + \left(\frac{\alpha_s}{2 \pi} \right)^2 A_j^{(2)}
 + \mathcal{O}(\alpha_s^3)\,, \nonumber \\
 E_j^{\lambda_1 \lambda_2} &= 
 \left(\frac{\alpha_s}{2 \pi} \right) E_j^{(1),\lambda_1 \lambda_2} 
 + \left(\frac{\alpha_s}{2 \pi} \right)^2 E_j^{(2),\lambda_1 \lambda_2}\,
 + \mathcal{O}(\alpha_s^3)\,.
\end{align}

\section{Calculation of the form factors}
\label{sec:calc}

The calculation of the coefficients $E_j^{\lambda_1 \lambda_2}$ proceeds as follows.
We produce all one- and two-loop Feynman diagrams relevant for $gg \to V_1 V_2$
using \qgraf~\cite{Nogueira:1991ex}. In particular we focus only on diagrams in classes 
$A$ and $B$ with massless quarks, for which we find $8$ diagrams at one loop and 
$138$ diagrams at two loops. Diagrams in class $F_V$, in fact, are simple three-point functions,
which sum up to zero due to charge-parity invariance.
The coefficients $A_j$ are then
calculated by applying the projectors defined in~\eqref{proj} on the different
Feynman diagrams. 
We insert the Feynman rules in our diagrams, where we employ the
Feynman-'t\;Hooft gauge ($\xi = 1$) for internal gluons.
After evaluation of Dirac traces and contraction of Lorentz indices
every Feynman diagram is expressed as linear combination of a large
number of scalar integrals. The latter belong to the family of the massless
four-point functions with two off-shell legs of different virtualities and
can be reduced to a small set of master integrals using 
integration-by-parts identities~\cite{Tkachov:1981wb,Chetyrkin:1981qh,Gehrmann:1999as,Laporta:2001dd}. 
We employ \reduze~2~\cite{vonManteuffel:2012np,Studerus:2009ye,Bauer:2000cp,fermat} to map 
all scalar integrals to the three integral families given in~\cite{Gehrmann:2014bfa}
and their crossed versions,
and subsequently to reduce them to master integrals.
In this way, we obtain analytical expressions for the coefficients $A_j$ as linear combinations of 
the latter. For the master integrals we employ the solutions presented in \cite{Gehrmann:2015ora}.
With the explicit expressions for the coefficients $A_j$ at the
different perturbative orders, it is easy to obtain the corresponding 
results for the $E_j^{\lambda_1 \lambda_2}$ using the formulae given 
in Appendix~\ref{sec:form}. \form~\cite{Vermaseren:2000nd} was used
extensively for all intermediate algebraic manipulations.\newline

Because of the lack of any tree-level contribution to the process $gg \to V_1V_2$,
the UV and IR pole-structure of the one- and two-loop amplitudes is very simple.
Clearly, the one-loop amplitude must be both UV- and IR-finite, and therefore
the pole structure of the two-loop amplitude will be, \textsl{effectively},
what one usually encounters for a one-loop QCD amplitude.
As discussed above, QCD radiative corrections affect only the partonic amplitude
and can be taken into account via the $20$ independent
scalar coefficients $A_j$, see Eq.~\eqref{tensordec}.
Working in conventional dimensional regularisation, we may alternatively
consider the $18$ physically relevant form factors $E_i^{\lambda_1 \lambda_2}$ defined
as $d$ dimensional linear combinations of the $A_j$, see Appendix~\ref{sec:form}.
In what follows, all considerations regarding UV-renormalisation and the structure of
the IR poles of the partonic amplitude hold identically for any $A_j$ and $E^{\lambda_1 \lambda_2}_i$.
We will therefore focus on the scalar coefficients rather than on the full partonic amplitude, 
and use the symbol $\Omega$ to refer to any of the latter,
$$\Omega \in \left\{ A_j\,, E^{\lambda_1 \lambda_2}_i \right\}\,,\qquad \mbox{for any} 
\quad j=1,...,20, \quad i=1,...,9, \quad \lambda_1\lambda_2 = LL,LR\,.$$

We start by performing UV-renormalisation in the $\MS$ scheme. 
In massless QCD this amounts to replacing the bare coupling, $\alpha_0$,
with the renormalised one, 
$\alpha_s = \alpha_s(\mu^2)$, where $\mu$ is the renormalisation scale.
Here we only need the one-loop relation
\begin{equation}
\alpha_0\, \mu_0^{2 \epsilon}\, S_\epsilon = \alpha_s\, \mu^{2 \epsilon}
\left [ 1 - \frac{\beta_0}{\epsilon} \left( \frac{\alpha_s}{2 \pi} \right)+ 
\mathcal{O}(\alpha_s^2)\right]\,, 
\end{equation}
where
\begin{equation}
S_\epsilon = (4 \pi)^\epsilon \, \e^{-\epsilon \gamma}\,, 
\qquad \mbox{with the Euler-Mascheroni constant} \quad \gamma = 0.5772...\,,
\end{equation}
 $\epsilon = (4-d)/2$, $\mu_0$ is the mass-parameter introduced in dimensional regularisation to maintain 
a dimensionless coupling in the bare QCD Lagrangian density, and finally $\beta_0$
is the first order of the QCD $\beta$-function
\begin{equation}
\label{beta0}
\beta_0 = \frac{11 \,C_A - 4 \,T_F\,N_f}{6}\,, \quad \mbox{with}\quad C_A = N\,, 
\quad C_F = \frac{N^2-1}{2\,N}\,, \quad
T_F = \frac{1}{2}\,.  
\end{equation}
The renormalisation is performed at $\mu^2 = s$, the invariant mass squared of the vector-boson
pair. The renormalised form factors read then, in terms of the un-renormalised ones,
\begin{eqnarray}
\Omega^{(1)}  &=& 
S_\epsilon^{-1} \Omega^{(1),{\rm un}} ,  \nonumber \\
\Omega^{(2)} &=& 
S_\epsilon^{-2} \Omega^{(2),{\rm un}}  
-\frac{\beta_0 }{\epsilon} S_\epsilon^{-1}
\Omega^{(1),{\rm un}}\;.
\end{eqnarray}

After UV renormalisation, the two-loop  coefficients $\Omega^{(2)}$ 
contain still residual IR singularities. 
In any IR-safe observable these divergences are cancelled
by the corresponding ones produced in one-loop radiative processes with one more
external parton.
In the present case of $gg \to V_1 V_2$, as discussed already above,
the IR-poles at two loops are of NLO type and their structure has been know for a long time.
Here, we choose to follow the conventions used for the NNLO corrections
to $q\bar{q}\to V_1 V_2$ in~\cite{Gehrmann:2015ora}, which required a NNLO subtraction scheme.
The exact structure of the IR poles up to NNLO in QCD was predicted
first by Catani~\cite{Catani:1998bh}.
We present our results in a slightly modified scheme described in~\cite{Catani:2013tia},
which is well suited for the $q_T$-subtraction formalism. 

We define the IR finite amplitudes at renormalisation scale $\mu$
in terms of the UV renormalised ones as follows
\begin{align}
\Omega^{(1),\finite}_{q_T} &= \Omega^{(1)} \,,\nonumber \\
\Omega^{(2),\finite}_{q_T} &= \Omega^{(2)} - I_1(\epsilon)\, \Omega^{(1)} \,, \label{IRqT}
\\
\intertext{where for the gluon-fusion channel we have}
I_1(\epsilon) &= I_1^{soft}(\epsilon) + I_1^{coll}(\epsilon) \,,
\\[0.5ex]
I_1^{soft}(\epsilon) &= -\frac{\e^{\epsilon \gamma}}{\Gamma(1-\epsilon)} \left( \frac{\mu^2}{s}\right)^{\epsilon}\, 
\left( \frac{1}{\epsilon^2} + \frac{i \pi}{\epsilon} + \delta_{q_T}^{(0)} \right)\,C_A
\,,\\ 
I_1^{coll}(\epsilon) &=- \frac{1}{\epsilon} \beta_0\, \left( \frac{\mu^2}{s}\right)^{\epsilon}\,.
\end{align}
Following~\cite{Catani:2013tia} we then put
$\delta_{q_T}^{(0)} = 0$.     
We provide the explicit analytical results for the finite remainders of the coefficients 
$A_j$ in this scheme, obtained for $\mu^2 = s$, on our project page at \hepforge.

Finally, it is straight-forward to convert these finite remainders 
into the Catani's original subtraction scheme~\cite{Catani:1998bh}, as extensively described in~\cite{Gehrmann:2015ora}.
For the present case we obtain the conversion formulae
\begin{align}\label{qt2catani}
\Omega^{(1),\finite}_{\text{Catani}} &= \Omega^{(1),\finite}_{q_T} ,\nonumber\\
\Omega^{(2),\finite}_{\text{Catani}} &= \Omega^{(2),\finite}_{q_T} + \Delta I_1 \, \Omega^{(1),\finite}_{q_T},
\end{align}
with $\Delta I_1$, in the case of a $gg$ initial state, is given by 
\begin{align}\label{qt2catanicoeff}
 \Delta I_1 &=   -\frac{1}{2} \pi^2 C_A + i \pi \beta_0 \,.
\end{align}

In order to test the correctness of our results we have performed a number of checks,
which we list in the following.
\begin{enumerate}
\item First of all, we computed explicitly all one- and two-loop diagrams relevant for $gg \to V_1 V_2$,
including those diagrams in class $B$ which are expected not to give any contribution due to Furry's theorem,
see Section~\ref{sec:calc}. We have verified that, after reduction to master integrals, all diagrams in 
class $B$ sum up to zero.

\item We have verified explicitly that the coefficients $A_j$ respect the expected symmetry relations 
derived in~\eqref{Ajx12} and \eqref{Ajx34}.

\item We have verified explicitly that the IR poles of the two-loop amplitude have the structure 
predicted by Catani's formula, see Section~\ref{sec:calc}. This provides a strong
check of the correctness of the result.

\item We have performed a thorough comparison of our results with 
an independent calculation of the same process~\cite{Caola:2015ila}.
Specifically, we compared our results prior to UV renormalisation and
IR subtraction. While the representation of the amplitudes in terms of
spinor structures in~\cite{Caola:2015ila} has a different form than
our decomposition~\eqref{MllLL},
we found that both are equivalent.
For the full helicity amplitudes we have found perfect numerical agreement
at one- and two-loop order.
Moreover, expressing the form factors defined in~\cite{Caola:2015ila} as linear
combinations of our form factors $E_j^{\lambda_1 \lambda_2}$, we have verified
that for each of them independently we have perfect numerical agreement
at one- and two-loop order.
\end{enumerate}

\section{Numerical {\tt C++} implementation and results}
\label{sec:num}

\begin{figure}
\includegraphics[width=\textwidth]{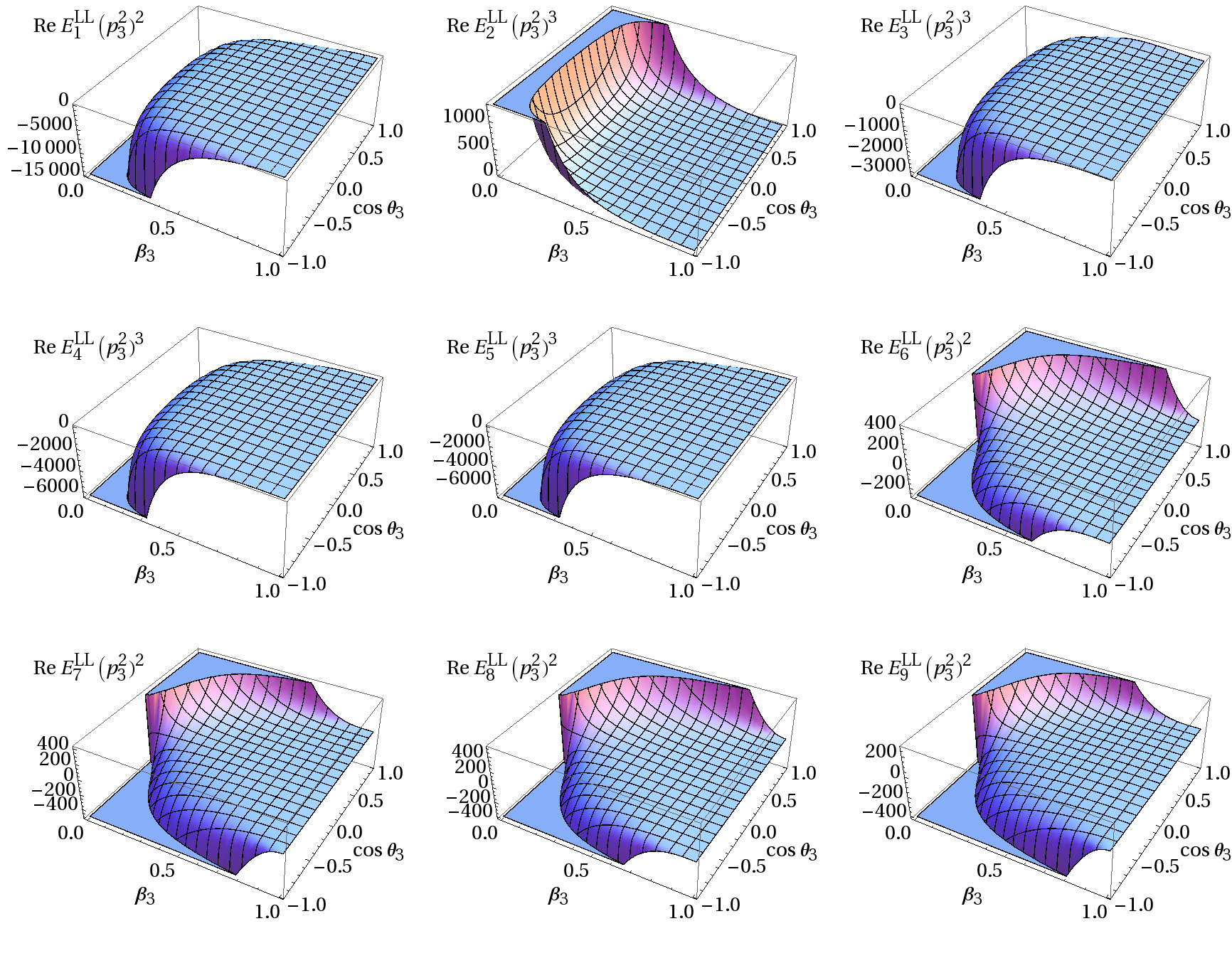}
\caption{\label{fig:plot3dejll}
Real parts of the two loop form factors $E_j^{(2),LL\,[A]}$
for the process $gg\to V_1V_2$.
The plots illustrate their dependence on the velocity, $\beta_3$,
and the cosine of the scattering angle, $\cos\theta_3$,
of the vector boson $V_1$,
where $p_4^2 = 2 p_3^2$ is chosen for the vector boson virtualities.
}
\end{figure}

\begin{figure}
\includegraphics[width=\textwidth]{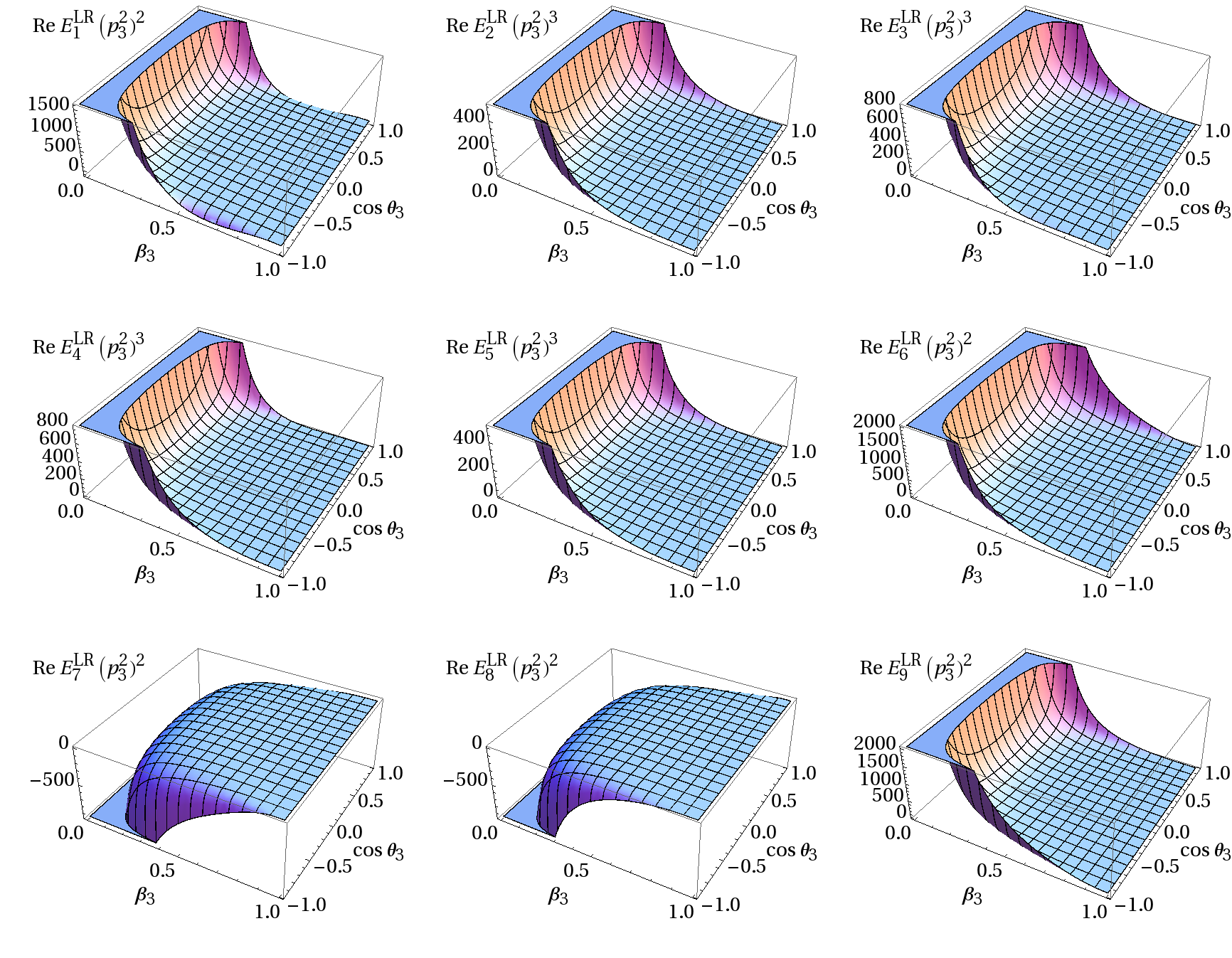}
\caption{\label{fig:plot3dejlr}
Real parts of the two loop form factors $E_j^{(2),LR\,[A]}$
for the process $gg\to V_1V_2$.
The plots illustrate their dependence on the velocity, $\beta_3$,
and the cosine of the scattering angle, $\cos\theta_3$,
of the vector boson $V_1$,
where $p_4^2 = 2 p_3^2$ is chosen for the vector boson virtualities.
}
\end{figure}

For the numerical evaluation of the helicity amplitudes for $gg \to V_1 V_2\to 4~\mathrm{leptons}$,
we implemented our results for the form factors $E_j^{\lambda_1 \lambda_2\,[A]}$
and $A_j^{[A]}$ at one- and two-loop order in a dedicated {\tt C++} code.
The implementation is based on the solutions for the master integrals presented
in~\cite{Gehrmann:2015ora}, which were specifically constructed for fast and reliable numerical
evaluations.
We organised our form factor implementation in form of a library, which is supplemented by a
simple command line interface.
We provide the software package for public download on \hepforge\ at
\url{http://vvamp.hepforge.org}.

For the numerical evaluation of the multiple polylogarithms encountered in the solutions for
the master integrals, we employ their implementation~\cite{Vollinga:2004sn} in the {\tt GiNaC}~\cite{Bauer:2000cp} library.
To identify and account for possible numerical instabilities of the form factors in collinear
or other potentially problematic regions of phase space, the code compares numerical evaluations,
which are obtained using different floating point data types, similar to the setup used in~\cite{Gehrmann:2015ora}.
If the results obtained with different precision settings differ beyond a user-defined
tolerance, the code successively increases the precision until the target precision is met.

For the rather central benchmark point of~\cite{Caola:2014iua}, the double precision mode of our code takes
roughly $600ms$ on a single computer core and results in at least $11$ significant digits for all of the
$E_j^{\lambda_1 \lambda_2\,[A]}$.
In order to estimate the actual precision, the default behaviour of our code
is to reevaluate the algebraic expressions in quad precision, which results in a total
run-time of roughly $3 s$ for this phase space point.
The run-time can increase further for regions close to the phase space
boundaries, where the multiple polylogarithms take more time to evaluate
and the precision control of our code may switch to higher precision
computations in order to return reliable numbers.
Depending on the precision setting and on the region of phase space, the
evaluations of the multiple polylogarithms and of the algebraic coefficients
may require comparable portions of the run-time.
However, in double precision mode or close to the phase space boundaries,
the run-time is dominated by multiple polylogarithm evaluations.
The described version of our code implements a minimal set of $9$ coefficients
$A_j$ and employs four evaluations of them with different kinematics
in order to derive the remaining form factors using crossing relations.
If required, it is straight-forward to further improve the evaluation
speed, either by proper caching of multiple polylogarithms or, at the
price of an increased code size, by an explicit implementation of all
form factors, as we did for the process $q \bar{q}' \to V_1 V_2$ in~\cite{Gehrmann:2015ora}.

In order to illustrate the form factors and the reliability of the code,
we used the latter to plot the real part of the two-loop form factors
for the case $p_4^2 = 2 p_3^2$ in Figures~\ref{fig:plot3dejll} and \ref{fig:plot3dejlr}.
In the plots, we vary the relativistic velocity $\beta_3$ and the cosine of
the scattering angle $\cos\theta_3$ of the vector boson $V_1$,
where $\beta_3= \kappa/(s+p_3^2-p_4^2)$ and $\cos\theta_3= (2t+s-p_3^2-p_4^2)/\kappa$.
Compared to the results for the form factors $E_j$ in the process
$q\bar{q}' \to V_1V_2$ in~\cite{Gehrmann:2015ora}, we observe
strong enhancements for the forward, backward and production threshold
regions for the form factors in the present case.
However, for the physical helicity amplitudes~\eqref{MllLL} we wish to point
out that an additional dampening (very) close to the aforementioned
phase space boundaries should be taken into account due to the additional overall
factors $C_{LL}$ and $C_{LR}$~\eqref{cll}.

\section{Conclusions}
\label{sec:conc}

In this paper we computed the two-loop massless QCD corrections 
to the helicity amplitudes for the production of pairs of off-shell
electroweak gauge bosons, $V_1 V_2$, in the gluon fusion channel.
For the calculation we employed the solutions for the master integrals
presented in~\cite{Gehrmann:2015ora}.
Contracting the diboson amplitude with the leptonic 
decay currents we have constructed the helicity amplitudes for
$gg \to V_1 V_2 \to 4~\mathrm{leptons}$. 
We have compared our results to an independent calculation~\cite{Caola:2015ila} and find
perfect agreement. 
Our results for these amplitudes provide the fundamental ingredient
required to compute the NLO corrections to diboson production processes in gluon fusion.
These corrections would contribute formally at N$^3$LO to the processes
$pp \to V_1V_2 + X$, but their inclusion may be important to match the 
expected experimental accuracy due to the large gluon luminosity at the LHC.
In particular studying their impact is required to obtain a more reliable estimate
of the theory uncertainty and to establish more precise constraints on the total Higgs decay 
width~\cite{Campbell:2011cu,Caola:2013yja,Campbell:2013una}.
We provide both analytical results and a {\tt C++} code for the numerical evaluation 
of the amplitudes on \hepforge\ at \url{http://vvamp.hepforge.org}.

\section*{Acknowledgements}
We are grateful to K.~Melnikov and F.~Caola for the comparison of our results with
the results of their independent calculation prior to
publication~\cite{Caola:2015ila}. We wish to thank K.~Melnikov and T.~Gehrmann
for clarifying discussions on different aspects of the calculation,
for interesting comments on the manuscript,
and for the encouragement to carry out this calculation to the end.
We thank the \hepforge\ team for providing web space for our project.
The Feynman graphs in this article have been drawn with
\jaxodraw~\cite{Binosi:2003yf,Vermaseren:1994je}.

\appendix

\section{Form factor relations}
\label{sec:form}

In this Appendix we present the explicit formulae needed in order to compute the 
$18$ form factors $E_j^{\lambda_1 \lambda_2}$ defined for the amplitude~\eqref{MllLL},
starting from the $20$ form factors $A_j$ defined in~\eqref{tensordec}.
For the $M_{LLLL}$ amplitude we find\\[-7pt]
\begin{align}
 &E_1^{LL} = \frac{2 A_1 + A_2 + A_3}{t\,u - p_3^2\,p_4^2} - \frac{A_{16}}{s}\,,\nonumber \\
& \nonumber \\[-7pt]
 &E_2^{LL} = \frac{A_{14}(t-p_4^2) - A_{12}(u-p_3^2) -s\,A_4}{s(t\,u - p_3^2\,p_4^2)} 
              + \frac{A_{17}}{2\,s}\,,\nonumber \\
& \nonumber \\[-7pt]
 &E_3^{LL} = \frac{A_{14}(u-p_4^2) - A_{13}(u-p_3^2) +A_2 +A_3 -s\,A_5}{s(t\,u - p_3^2\,p_4^2)} 
              + \frac{A_{18}}{2\,s}\,,\nonumber \\
& \nonumber \\[-7pt]
 &E_4^{LL} = \frac{A_{15}(t-p_4^2) - A_{12}(t-p_3^2) +A_2 +A_3 -s\,A_6}{s(t\,u - p_3^2\,p_4^2)} 
              + \frac{A_{19}}{2\,s}\,,\nonumber \\
& \nonumber \\[-7pt]
 &E_5^{LL} = \frac{A_{15}(u-p_4^2) - A_{13}(t-p_3^2) -s\,A_7}{s(t\,u - p_3^2\,p_4^2)} 
              + \frac{A_{20}}{2\,s} \,,\nonumber \\
& \nonumber \\[-7pt]
 &E_6^{LL} = \frac{(u-p_3^2) ( A_2- A_3 )}{s(t\,u - p_3^2\,p_4^2)} +\frac{A_{10}-A_{14}}{s}\,,
 \qquad\quad E_7^{LL} = \frac{(t-p_4^2) ( A_2- A_3 )}{s(t\,u - p_3^2\,p_4^2)} +\frac{A_{8}-A_{12}}{s}\,,\nonumber \\
& \nonumber \\[-7pt]
 &E_8^{LL} = \frac{(u-p_4^2) ( A_2- A_3 )}{s(t\,u - p_3^2\,p_4^2)} +\frac{A_{9}-A_{13}}{s}\,,
 \qquad\quad\; E_9^{LL} = \frac{(t-p_3^2) ( A_2- A_3 )}{s(t\,u - p_3^2\,p_4^2)} +\frac{A_{11}-A_{15}}{s}\,.
 \end{align}
For the $M_{LRLL}$ amplitude we have instead\\[-7pt]
\begin{align}
 E_1^{LR} &=  \frac{A_{2}+A_{3}}{t\,u - p_3^2\,p_4^2} + \frac{A_{16}}{s}\,, 
  \quad& E_2^{LR} &= - \frac{A_{17}}{2\,s}\,,\nonumber \\
 \nonumber \\[-7pt]
 E_3^{LR} &= \frac{A_2 +A_3}{s(t\,u - p_3^2\,p_4^2)} - \frac{A_{18}}{2\,s} \,,
 & E_4^{LR} &= \frac{A_2 +A_3}{s(t\,u - p_3^2\,p_4^2)} - \frac{A_{19}}{2\,s}\,,\nonumber \\
 \nonumber \\[-7pt]
 E_5^{LR} &=  - \frac{A_{20}}{2\,s} \,, 
 & E_6^{LR} &= \frac{(u-p_3^2)(A_2 +A_3)}{s(t\,u - p_3^2\,p_4^2)} - \frac{A_{10}+A_{14}}{s} \,,\nonumber \\
 \nonumber \\[-7pt]
 E_7^{LR} &= - \frac{(t-p_4^2)(A_2 +A_3)}{s(t\,u - p_3^2\,p_4^2)} - \frac{A_{8}+A_{12}}{s} \,,
 & E_8^{LR} &= - \frac{(u-p_4^2)(A_2 +A_3)}{s(t\,u - p_3^2\,p_4^2)} - \frac{A_{9}+A_{13}}{s} \,,\nonumber \\
 \nonumber \\[-7pt]
 E_9^{LR} &= \frac{(t-p_3^2)(A_2 +A_3)}{s(t\,u - p_3^2\,p_4^2)} - \frac{A_{11}+A_{15}}{s}\,.
 &&
 \end{align}

Let us consider the behaviour of the form factors $E_j^{\lambda_1 \lambda_2}$
under the two permutations $\pi_{12}$ and $\pi_{34}$ defined in~\eqref{perm}.
Using the crossing relations for the form factors $A_j^{[A]}$~\eqref{Ajx12} and \eqref{Ajx34} one easily finds
the corresponding ones for the form factors $E_j^{\lambda_1 \lambda_2\,{[A]}}$.
To simplify our notation, we drop the superscript $[A]$ in the following.
Under permutation $\pi_{12}$ we obtain
\begin{align}
E^{LL}_1(s,u,p_3^2,p_4^2) &= E^{LL}_1(s,t,p_3^2,p_4^2)\,, \nonumber \\
E^{LL}_2(s,u,p_3^2,p_4^2) &= E^{LL}_5(s,t,p_3^2,p_4^2) 
+ \frac{(u - p_4^2) E^{LL}_9(s,t,p_3^2,p_4^2) 
-  (t -p_3^2 ) E^{LL}_8(s,t,p_3^2,p_4^2)}{t u - p_3^2\,p_4^2} \,,\nonumber \\
E^{LL}_3(s,u,p_3^2,p_4^2) &= E^{LL}_4(s,t,p_3^2,p_4^2) 
+ \frac{(t - p_4^2) E^{LL}_9(s,t,p_3^2,p_4^2) 
-  (t -p_3^2 ) E^{LL}_7(s,t,p_3^2,p_4^2)}{t u - p_3^2\,p_4^2}\,,\nonumber \\
E^{LL}_6(s,u,p_3^2,p_4^2) &= - E^{LL}_9(s,t,p_3^2,p_4^2)\,, \qquad\;\;\;
E^{LL}_7(s,u,p_3^2,p_4^2) = - E^{LL}_8(s,t,p_3^2,p_4^2)
\end{align}
and
\begin{alignat}{3}
E^{LR}_1(s,u,p_3^2,p_4^2) &= E^{LR}_1(s,t,p_3^2,p_4^2)\,,\qquad\;\;\,\,
&E^{LR}_2(s,u,p_3^2,p_4^2) &= E^{LR}_5(s,t,p_3^2,p_4^2)\,,\qquad\phantom{=}\nonumber \\
E^{LR}_3(s,u,p_3^2,p_4^2) &= E^{LR}_4(s,t,p_3^2,p_4^2)\,,
&E^{LR}_6(s,u,p_3^2,p_4^2) &= E^{LR}_9(s,t,p_3^2,p_4^2)\,,\nonumber \\
E^{LR}_7(s,u,p_3^2,p_4^2) &= E^{LR}_8(s,t,p_3^2,p_4^2)\,.&
&
\end{alignat}
Under permutation $\pi_{34}$, instead, the form factors for both the $LL$ and $LR$
helicity configurations transform in the same way
\begin{alignat}{3}
E^{\lambda_1 \lambda_2}_1(s,u,p_4^2,p_3^2) &= E^{\lambda_1 \lambda_2}_1(s,t,p_3^2,p_4^2)\,,\quad\;\; 
&E^{\lambda_1 \lambda_2}_2(s,u,p_4^2,p_3^2) &= E^{\lambda_1 \lambda_2}_2(s,t,p_3^2,p_4^2)\,,\nonumber \\
E^{\lambda_1 \lambda_2}_3(s,u,p_4^2,p_3^2) &= E^{\lambda_1 \lambda_2}_4(s,t,p_3^2,p_4^2)\,,
&E^{\lambda_1 \lambda_2}_5(s,u,p_4^2,p_3^2) &= E^{\lambda_1 \lambda_2}_5(s,t,p_3^2,p_4^2)\,,\nonumber \\
E^{\lambda_1 \lambda_2}_6(s,u,p_4^2,p_3^2) &= -E^{\lambda_1 \lambda_2}_7(s,t,p_3^2,p_4^2)\,,
&E^{\lambda_1 \lambda_2}_8(s,u,p_4^2,p_3^2)& = -E^{\lambda_1 \lambda_2}_9(s,t,p_3^2,p_4^2)\,.
\end{alignat}
Exploiting all of these crossing relations we find that only $9$ out of
the $18$ form factors $E_j^{\lambda_1\lambda_2}$ are effectively independent,
while the other $9$ can be obtained by the crossing rules above.
The number of independent form factors $E_j^{\lambda_1\lambda_2}$  coincides
with the number of independent form factors $A_j$ found in Section~\ref{sec:tens}.

\bibliographystyle{JHEPvtag}   
\bibliography{Biblio}     

\end{document}